\documentclass[prl, twocolumn,superscriptaddress,preprintnumbers,a4paper,amsmath,amssymb,showpacs,floatfix]{revtex4}

\usepackage{graphicx}
\usepackage{color}\color[rgb]{0.000,0.000,0.000} %used for font color
\usepackage{amssymb} %maths
\usepackage{amsmath} %maths
\usepackage{multirow}%%%%new bit
\usepackage{nicefrac}

\begin{document}
\newcommand{\mn}{MnS$_{2}$}
\newcommand{\te}{$_{t}$}
\newcommand{\ot}{$_{o}$}
\newcommand{\Tc}{T$_{C}$}
\newcommand{\Ts}{T$_{s}$}
\newcommand{\Tn}{$T_\mathrm{N}$}
\newcommand{\MuB}{$\mu_\mathrm{B}$}
\newcommand{\cu}{CuMnO$_{2}$}
\newcommand{\na}{NaMnO$_{2}$}

%\preprint{}
\title{Spin-chain correlations in the frustrated triangular lattice material CuMnO$_{2}$}

\author{Simon A. J. Kimber}
\email[Email of corresponding author:]{simon.kimber@u-bourgogne.fr}
\affiliation{Universit\'e Bourgogne-Franche Comt\'e, Universit\'e de Bourgogne, ICB-Laboratoire Interdisciplinaire Carnot de Bourgogne, B\^atiment Sciences Mirande, 9 Avenue Alain Savary, B-P. 47870, 21078 Dijon Cedex, France.}

\author{Andrew Wildes}
\affiliation{Institut Laue Langevin (ILL), 6 rue Jules Horowitz, BP 220, 38043  Grenoble Cedex 9, France.}

\author{Hannu Mutka}
\affiliation{Institut Laue Langevin (ILL), 6 rue Jules Horowitz, BP 220, 38043  Grenoble Cedex 9, France.}

\author{Jan-Willem G. Bos}
\affiliation{Institute of Chemical Sciences and Centre for Advanced Energy Storage and Recovery, School of Engineering and Physical Sciences, Heriot-Watt University, Edinburgh, EH14 4AS, UK.}
\author{Dimitri N. Argyriou}
\affiliation{European Spallation Source ESS AB, Box 176, 22100, Lund, Sweden}

\date{\today}
%\preprint{}

\begin{abstract}
The Ising triangular lattice remains the classic test-case for frustrated magnetism. Here we report neutron scattering measurements of short range magnetic order in \cu, which consists of a distorted lattice of Mn$^{3+}$~spins with single-ion anisotropy. Physical property measurements on \cu~are consistent with 1D correlations caused by anisotropic orbital occupation. However the diffuse magnetic neutron scattering seen in powder measurements has previously been fitted by 2D Warren-type correlations. Using neutron spectroscopy, we show that paramagnetic fluctuations persist up to $\sim$25 meV above $T_{N}$= 65 K. This is comparable to the incident energy of typical diffractometers, and results in a smearing of the energy integrated signal, which hence cannot be analysed in the quasi-static approximation. We use low energy XYZ polarised neutron scattering to extract the purely magnetic (quasi)-static signal. This is fitted by reverse Monte Carlo analysis, which reveals that two directions in the triangular layers are perfectly frustrated in the classical spin-liquid phase at 75 K. Strong antiferromagnetic correlations are only found along the $b$-axis, and our results hence unify the pictures seen by neutron scattering and macroscopic physical property measurements. 
\end{abstract}
\maketitle

\section{\label{sec:level1}INTRODUCTION}
The delafossite family of triangular lattice ABO$_2$ materials (where A is Cu, Ag, Pd, and B is a transition metal) are a fertile source of interesting physics. The magnetic properties of these materials are highly influenced by the choice of transition metal, for example V$^3$$^+$ compounds show interesting orbital physics~\cite{NaVO2}~and Cr$^3$$^+$ compounds are excellent 2D Heisenberg or XY model magnets~\cite{frontzek2011magnetic}. Furthermore, the choice of B-site cation also sensitively controls the electronic and lattice properties. For example, materials with A = Pd are often metallic~\cite{takatsu2009critical}~and in closely related Na$_0$$_.$$_7$CoO$_2$, sodium ordering transitions strongly influence the transport properties of the CoO$_2$ layer~\cite{roger2007patterning}. 
Particularly strong coupling between magnetic and lattice degrees of freedom is found in delafossites with Ising single ion anisotropy, the best known example of which is  CuFeO$_2$. In this compound, a structural phase transition which breaks the rhombohedral parent symmetry precedes~\cite{mitsuda1998partially}~magnetic order at low temperature~\cite{ye2006spontaneous}. Further evidence for Ising character comes from the wealth of complex magnetic structures generated by applied magnetic fields, and gapped spin wave excitations. Recently, another family of Ising triangular lattice compounds related to the delafossites has attracted attention~\cite{giot2007magnetoelastic,zorko2008magnetic} . In \na~and \cu, an anisotropic triangular lattice is found at room temperature due to Mn$^{3+}$ orbital order (see Fig. 1a). The interaction along the short direction in the triangular lattice (J$_1$ in Fig. 1b) is strongly antiferromagnetic due to direct exchange, whereas the superexchange interactions between chains (J$_2$) are thought to be weaker as the bond angles ($\sim$97~$^\circ$ ) are close to the FM-AFM crossover region.  The most obvious experimental manifestation of this effect is the magnetic susceptibility, which shows a very large hump at $\sim$90 K attributed to low dimensional correlations \cite{damay2009spin}. The magnetic correlations in these materials hence show  lower dimensionality than the crystal structure. This is a familiar scenario in frustrated magnetic materials e.g. Cs$_{2}$CuCl$_{4}$, shows rather anisotropic 2D behaviour, despite the nearly symmetric Cu triangular lattice~\cite{coldea2003extended}.\\
Varying models have been used to interpret magnetic neutron scattering data on these materials. For \na, Stock $et~al$~used the single mode approximation with J$_{1}$=73 K to analyse powder inelastic data~\cite{stock2009one}. The 1D nature of the excitations was recently confirmed by single crystal measurements~\cite{dally2018amplitude}. In contrast, \cu~has been more commonly discussed in a 2D picture. For example, an intense asymmetric peak of magnetic origin has been reported above $T_{N}$~in at least three neutron powder diffraction experiments~\cite{damay2009spin,garlea2011tuning,terada2011magnetic}. This has been fitted using a so-called Warren function, and a 2D correlation length extracted. A 3 meV feature related to 2D spin-liquid fluctuations is also said to co-exist with magnetic order at low temperature~\cite{terada2011magnetic}. Very recently, the magnetic pair distribution function (mPDF) method has also been applied to both materials~\cite{frandsen2020nanoscale}. This method detected short-range magnetic order above $T_{N}$, however, the correlation lengths were modelled assuming 3D spherical domains.\\
\begin{figure}[tb!]
\begin{center}
\includegraphics[scale=0.5]{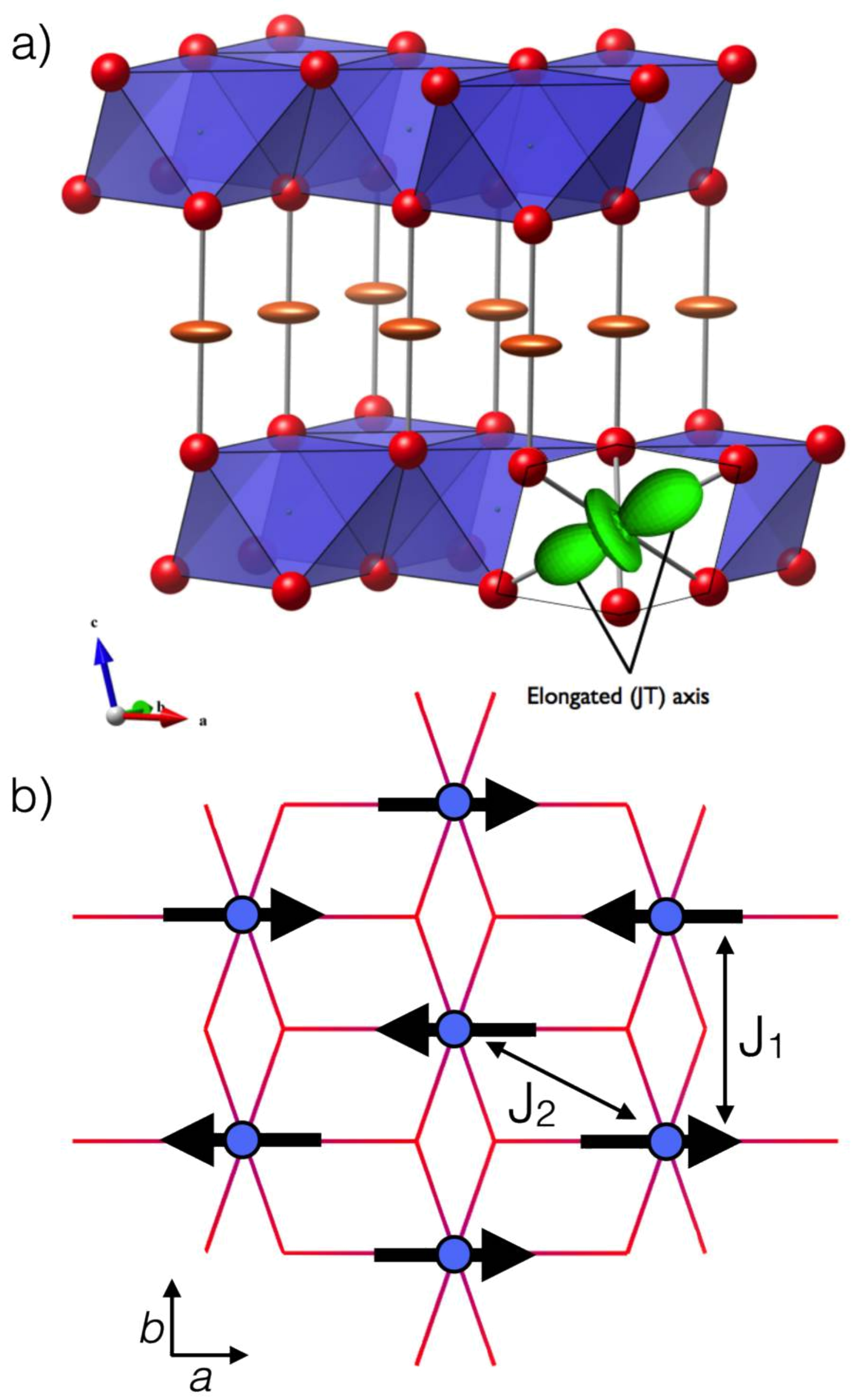}
\caption{(color online) (a) Crystal structure~\cite{momma2011vesta}~of monoclinic  \cu, the MnO$_{6}$~octahedra are shown in blue, and interlayer Cu$^{+}$~cations are shown ; (b) Principle magnetic exchange interactions in the triangular layers of monoclinic (paramagnetic) \cu. Blue spheres are Mn$^{3+}$ cations, with the magnetic structure found below 65 K shown with black arrows. Thin red lines show the Mn-O bonds.}
\label{Fig1}
\end{center}
\end{figure}
 As we discuss in more detail below, analysing diffuse scattering from energy integrated measurements requires that the quasi-static approximation holds. That is to say, the incident probe energy ($E_{i}$) must significantly exceed the bandwidth of the excitations ($\Delta$E). Here we directly measure the latter quantity using thermal neutron spectroscopy, before applying cold XYZ polarised diffuse scattering to measure the (quasi)-elastic infinite time spin-spin correlations. Our results reveal that: 1) A broad inelastic magnetic signal is found above $T_{N}$, which extends up to at least 25 meV; 2) This energy scale is comparable to the $E_{i}$~used for most energy integrated measurements in the literature, thus distorting the shape of the diffuse signal; 3) The elastic diffuse scattering signal is poorly fitted by 2D expressions, and we report reverse Monte Carlo simulations that confirm largely 1D spin correlations; 4) Our cold neutron measurements detect no evidence for a low energy ($<$ 3.47 meV) diffuse component that co-exists with the magnetic Bragg peaks. 

\section{\label{sec:level1}EXPERIMENTAL}
Our powder sample of \cu~was synthesised from Cu powder (Sigma Aldrich 99.999\%) and MnO$_{2}$ (Sigma Aldrich 99.99\%), which were pelleted and heated in a sealed, evacuated silica tube at 960 $^{\circ}$C for 3 x 12 hours with intermediate regrinds. Laboratory X-ray diffraction revealed a highly crystalline product with the $C2/m$~structure previously reported. We used a variety of angle-dispersive neutron scattering instrumentation to characterise the structure and dynamics. These were the powder diffractometer E9 at the (then) Hahn-Meitner Institute, Berlin, Germany, the time of flight spectrometer IN4c, and the diffuse polarised spectrometer~\cite{stewart2009disordered,fennell2017wavevector}~D7 (both at the Institut Laue Langevin, Grenoble, France). We detected no evidence for antisite disorder by Rietveld refinement against the E9 data, and refined cell parameters, coordinates etc were comparable to literature reports. In all cases, Orange He flow cryostats were used for temperature control, and the sample was held in a vanadium container on E9, and Al cans on IN4c and D7. A background measurement of the empty can was made and subtracted from the final data on the latter two instruments. Standard data reduction routines in \textsc{lamp}~were used for the IN4c and D7 experiments.\\
A neutron detector at fixed angle counts both elastically and inelastically scattered neutrons. However, the trajectory traced through momentum and energy transfer space,  S(Q,$\omega$) is curved. This can therefore distort an inelastic diffuse signal unless the incident energy, $E_{i}$ far exceeds the energy scale of the fluctuations responsible. On IN4, time of flight resolution is used to correct this effect. To calculate the S(Q,$\omega$) space integrated over by E9 and D7 (in diffraction mode),  we the reported detector bank angles and the kinematic condition below:\\
\begin{equation}
\frac{\hbar Q^{2}}{2m}=E_{i}+E_{f}-2\sqrt{E_{i}E_{f}}\cos 2 \theta
\end{equation}
Here $\theta$~is the detector angle, $E_i$~and $E_f$~are the incident/final energies, and $m$ is the mass of the neutron. The incident wavelengths (energies) used for the three instruments were: E9, 1.797 \AA~(25 meV); IN4c, 1.1 \AA~(68 meV); D7, 4.855 \AA~(3.47 meV). On E9, data were collected in 15 K steps from 1.5 to 300 K. At lower temperatures ($<$~67 K), our sample showed the same triclinic structural distortion and magnetic order as reported elsewhere ~\cite{damay2009spin,garlea2011tuning,terada2011magnetic}.  A strong diffuse magnetic feature is seen even at room temperature. In several other works, this was fitted using the Warren peak shape, which describes 2D correlations:\\
\begin{equation}
I_{hk}=s \left[\frac{\zeta}{\sqrt{\pi\lambda}}\right]^{1/2}\frac{M_{hk}\mid~F_{M}\mid^2F(a)}{(\sin\theta)^{3/2}}
\end{equation}
Here, $s$~is a scale factor, $\theta$~is the scattering angle, $\zeta$ is the 2D correlation length, $\lambda$~the wavelength, $M_{hk}$~ the reflection multiplicity, $F_{M}$ ~the magnetic form factor, and the function $F(a)$~is given by:\\
\begin{equation}
F(a)=\int^{\infty}_{0}e^{-(x^2-a)^2}dx
\end{equation}
where (with $\theta_B$~the Bragg angle of the reflection):
\begin{equation}
a=(2\zeta\sqrt{\pi}/\lambda)(\sin\theta-\sin\theta_B)
\end{equation}
A short Python code was written to refine the correlation length and scale parameter against the experimental data. For fits against the E9 data the (001) structural Bragg peak was manually excluded.\\
On D7, we operated in diffraction mode (i.e. integrating over all final energies). Using the standard XYZ method~\cite{ehlers2013generalization}, the scattered signal was separated into magnetic, nuclear, spin incoherent, and combined nuclear/isotope incoherent components. The data analysed here represent the average magnetic scattering cross section $(d\sigma/d\Omega)_{mag}$. We analysed the data from D7 using the SpinVert Reverse Monte Carlo code~\cite{paddison2013spinvert}. We assumed Ising anisotropy and the preferred axis was assumed to lie along the long direction of the Jahn-Teller distorted MnO$_{6}$~octahedra. Note that refinements using Heisenberg spins gave equally good fits, but radically different spin-spin correlations. These were rejected due to the large excitation gap we detected in the inelastic neutron scattering data reported below. The tabulated $j_{0}$~form factor for Mn$^{3+}$~was used, and a simulation box of 20 x 20 x 10 unit cells was used with 500 moves per spin and a weight of 10. Ten independent refinements were performed sequentially and averaged. The SpinDiff program was used to simulate experimental single crystal diffuse scattering~\cite{paddison2013spinvert}.\\

\section{\label{sec:level1}RESULTS AND DISCUSSIONS}
\begin{figure}[tb!]
\begin{center}
\includegraphics[scale=0.4]{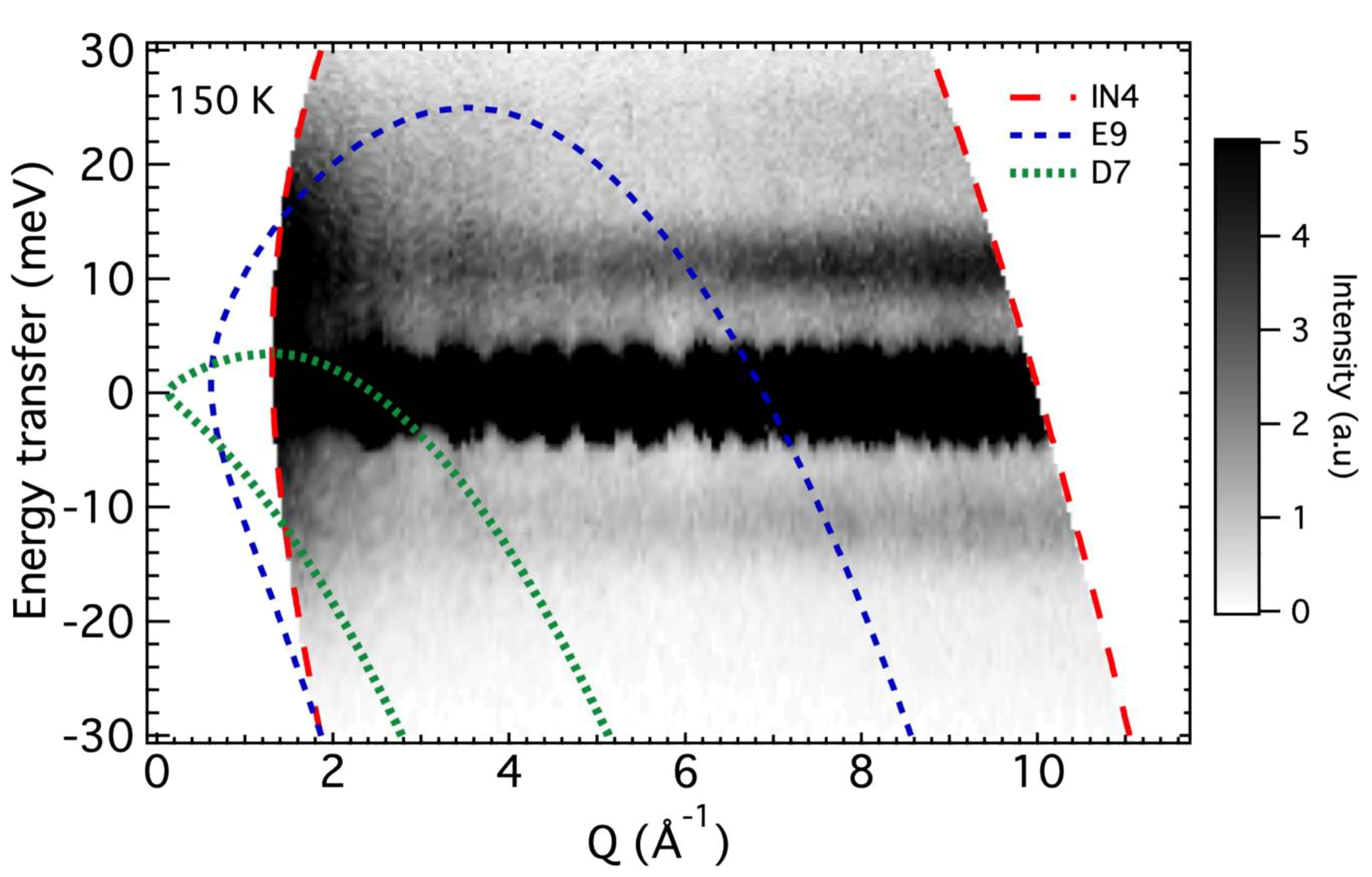}
\caption{(color online) Overview of the inelastic neutron scattering response of \cu~in the paramagnetic phase at 150 K. Data were collected on IN4c using an incident energy of 68 meV. The dashed lines show the area in S(Q,$\omega$) probed by the three instruments used here. The inelastic response contains two main features: 1) A low-Q plume of magnetic scattering and 2) dispersion-less phonon mode at ca. 12 meV.}
\label{Fig2}
\end{center}
\end{figure}
Figure two shows an overview of the inelastic response of \cu~in the paramagnetic phase at 150 K. Two clear features are seen: 1) A low-Q plume of magnetic scattering and 2) a phonon mode at ca. 12 meV. In this work we concentrate on the former, we will describe the origin of the latter (and its link to $c$-axis negative thermal expansion) elsewhere.  We have also superimposed the trajectories of the lowest and highest angle detectors on E9 and D7, as calculated for our energy integrated measurements using Eqn. 1. Clearly, the magnetic signal is only partially captured on E9 at this temperature, and the overall energy scale is similar to the $E_{i}$=25 meV used in the experiment. As justified below, the D7 measurements effectively probe only the (quasi)-static spin-spin correlations in our experiments. We next examine the magnetic scattering in more detail, by plotting the IN4c signal integrated over the range 1.25 $<$~Q $<$~2.25 \AA$^{-1}$~as shown in Fig.  3. At 150 K, this could be parameterised by a single damped harmonic oscillator\cite{faak1997phonon,krimmel2009spin}~using the equation below:\\
\begin{equation}
S(Q,\omega)=\frac{A_{\textrm{DHO}}\omega\Gamma_{\textrm{DHO}}}{(\omega^2-\omega^2_{\textrm{DHO}})+(\omega\Gamma_{\textrm{DHO}})^2}\cdot\frac{1}{1-e^{-\hbar\omega/k_BT}}
\end{equation} 
\begin{figure}[tb!]
\begin{center}
\includegraphics[scale=0.5]{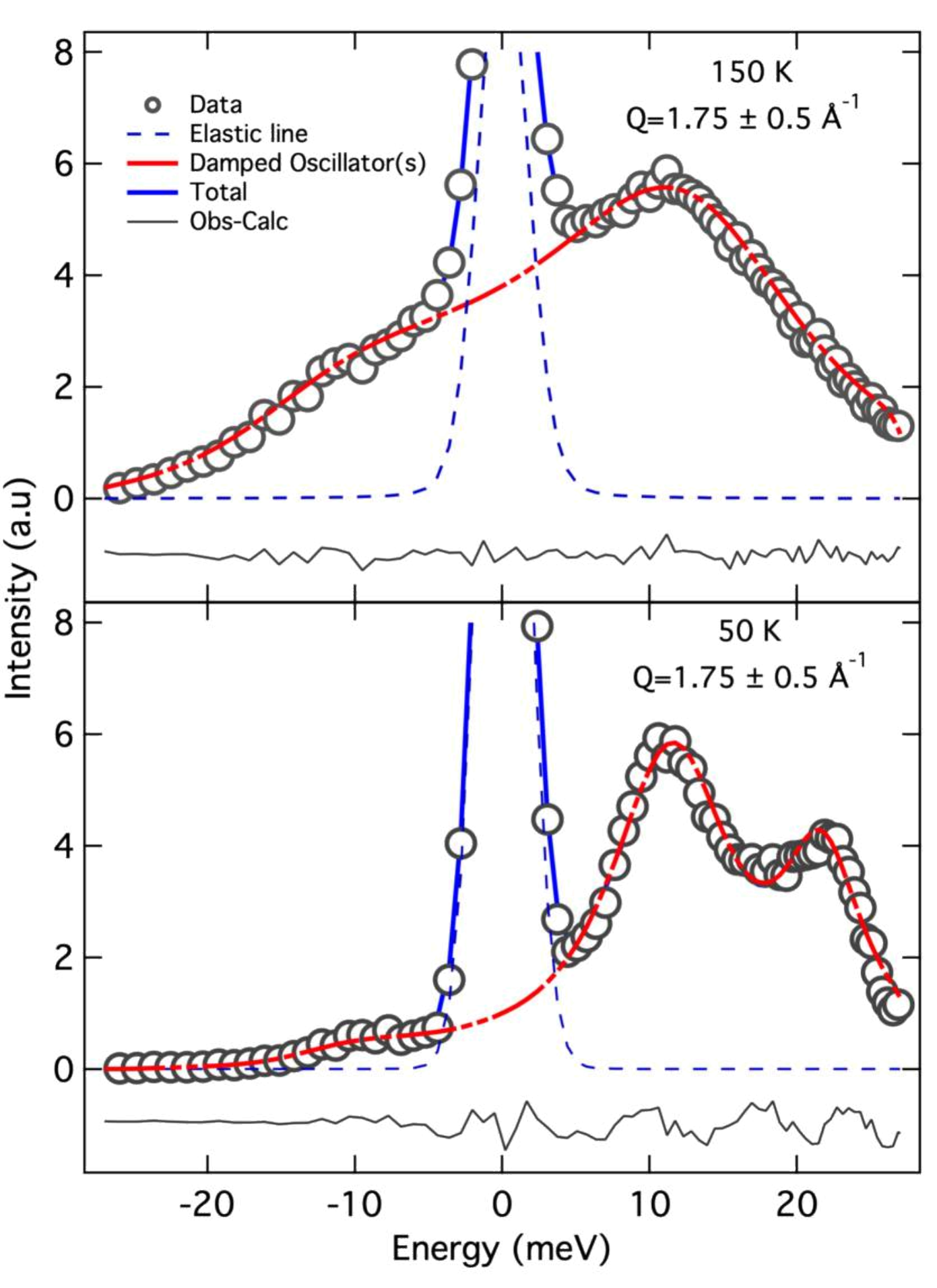}
\caption{(color online) (top) Fit of the damped oscillator model (Eqn. 5) to a cut through the elastic line of the IN4c data at 150 K. This model includes a Gaussian resolution function, a single oscillator and a quasi-elastic Lorentzian component; (bottom) Fit of the same cut just below $T_{N}$=65 K. Two oscillators are required to describe the data at the temperature. Error bars are smaller than the points.}
\label{Fig3}
\end{center}
\end{figure}
Here the amplitude of the mode is given by $A_{DHO}$, while the line width is $\Gamma_{DHO}$~and the position is given by $\omega$. The second term is the Bose factor, which relates the ratio of the neutron energy gain and energy loss intensities to the temperature ($k_{B}T$). In addition, we discovered a significant quasi-elastic broadening of the elastic line with a Lorentzian line shape. The fit shown is the result of convoluting these model expressions (and a Gaussian elastic line) with a Gaussian that represents the experimental energy resolution. This was determined by peak fitting an equivalent measurement on vanadium. At 150 K, the energy of the mode was $\omega$=17.2(1) meV. Below the magnetic ordering temperature, this features splits into two distinct spin-wave branches (at $\sim$13 and $\sim$23 meV), with a notable excitation gap (Fig. 3 bottom). The latter is consistent with a significant single-ion anisotropy. Neutron scattering measurements with low incident energy (e.g. on D7 with $E_{i}$=3.47 meV) are hence measuring only the (quasi)-elastic signal. The fate of the anti-Stokes scattering, which extends to higher energy transfer, is discussed in more detail below.\\
The results of our D7 experiment at 1.5 K (deep within the ordered phase) are shown in Fig. 4. We have plotted the total signal, as well as the nuclear and spin-flip magnetic components. The principal magnetic Bragg relections of \cu~are clearly observed over this Q-range, and our sample can be seen to show both the principle $k_{1}=(\frac{\bar{1}}{2},\frac{1}{2},\frac{1}{2})$~and $k_{2}=(\frac{\bar{1}}{2},\frac{1}{2},0)$~wavevectors previously reported~\cite{garlea2011tuning}. Peaks originating from the latter are extremely weak, as expected~\cite{garlea2011tuning} for near stoichiometric samples of \cu. Notable by its absence is any significant trace of a diffuse magnetic signal accompanying the Bragg peaks. We note that the kinematic window (Fig. 2) of D7 in this setting is peaked exactly where the magnetic signal is strongest ($\sim$Q=1.3 \AA$^{-1}$),as this Q-value is close to 2$\Theta$=90$^{\circ}$. This rules out static or inelastic disordered correlations with energy scales of up to $E_{i}$=3.47 meV. It is possible that the "2D spin liquid" signal previously observed in this region therefore reflects the resolution tails of the magnetic Bragg reflections~\cite{terada2011magnetic}.\\
The results of our (energy integrated) E9 powder diffraction measurements in the paramagnetic phase are consistent with earlier reports \cite{damay2009spin,garlea2011tuning,terada2011magnetic}. A strong asymmetric magnetic feature is seen to persist up to at least 300 K (already indicative of an energy scale of ca. 25 meV).This begins to rise around Q=1 \AA$^{-1}$, close to where the (001) Bragg reflection is found. Upon cooling to just above the antiferromagnetic ordering (N\'eel) temperature, this sharpens and grow in intensity. We fitted this feature to equation (2) at a range of temperatures, yielding a 2D correlation length of 20(1) \AA~ at 85 K (Fig. 5a). This parameter showed only a very weak temperature dependence (not shown) as reported elsewhere. On the face of it, this is unusual, and in fact, close examination of the data shows that the fit is actually fairly poor (Fig. 5a), especially on the high-Q side of the asymmetric feature, where a linear fit produces a lower residual. The origin of equation (2) is Warren's seminal work~\cite{warren1941x}~on disordered graphitic carbons. The asymmetric feature is the result of powder averaging rods of diffuse scattering from two dimensional order, which gives a well defined lower bound at low-Q and a long tail at higher-Q.  Obviously, for structural turbostratic disorder, one need not worry about inelastic effects. However, as discussed above, the quasi-static condition clearly does not hold for this measurement (see Fig. 2). The shape of the diffuse signal is thus distorted, and the assumption of 2D correlations wrong. 
\begin{figure}[tb!]
\begin{center}
\includegraphics[scale=0.55]{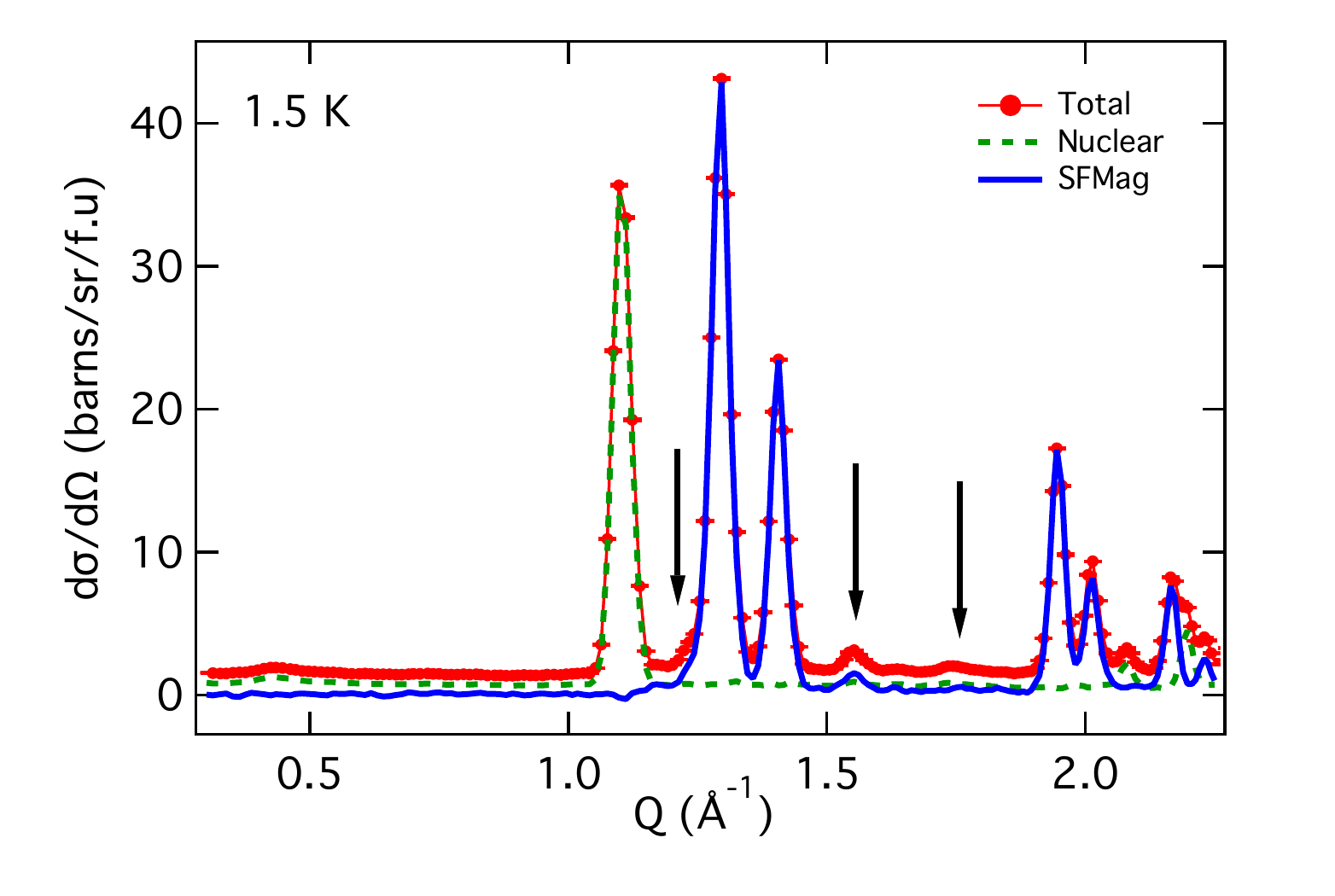}
\caption{(color online) Results of the D7 XYZ polarised neutron scattering experiment at 1.5 K. The data are plotted as the total signal, coherent nuclear and spin-flip magnetic parts. Peaks arising from the secondary $k_{2}=(\frac{\bar{1}}{2},\frac{1}{2},0)$~wavevector are marked with arrows. Note the absence of any significant diffuse contribution under the magnetic Bragg peaks.}
\label{Fig4}
\end{center}
\end{figure}
\begin{figure*}[tb!]
\begin{center}
\includegraphics[scale=0.5]{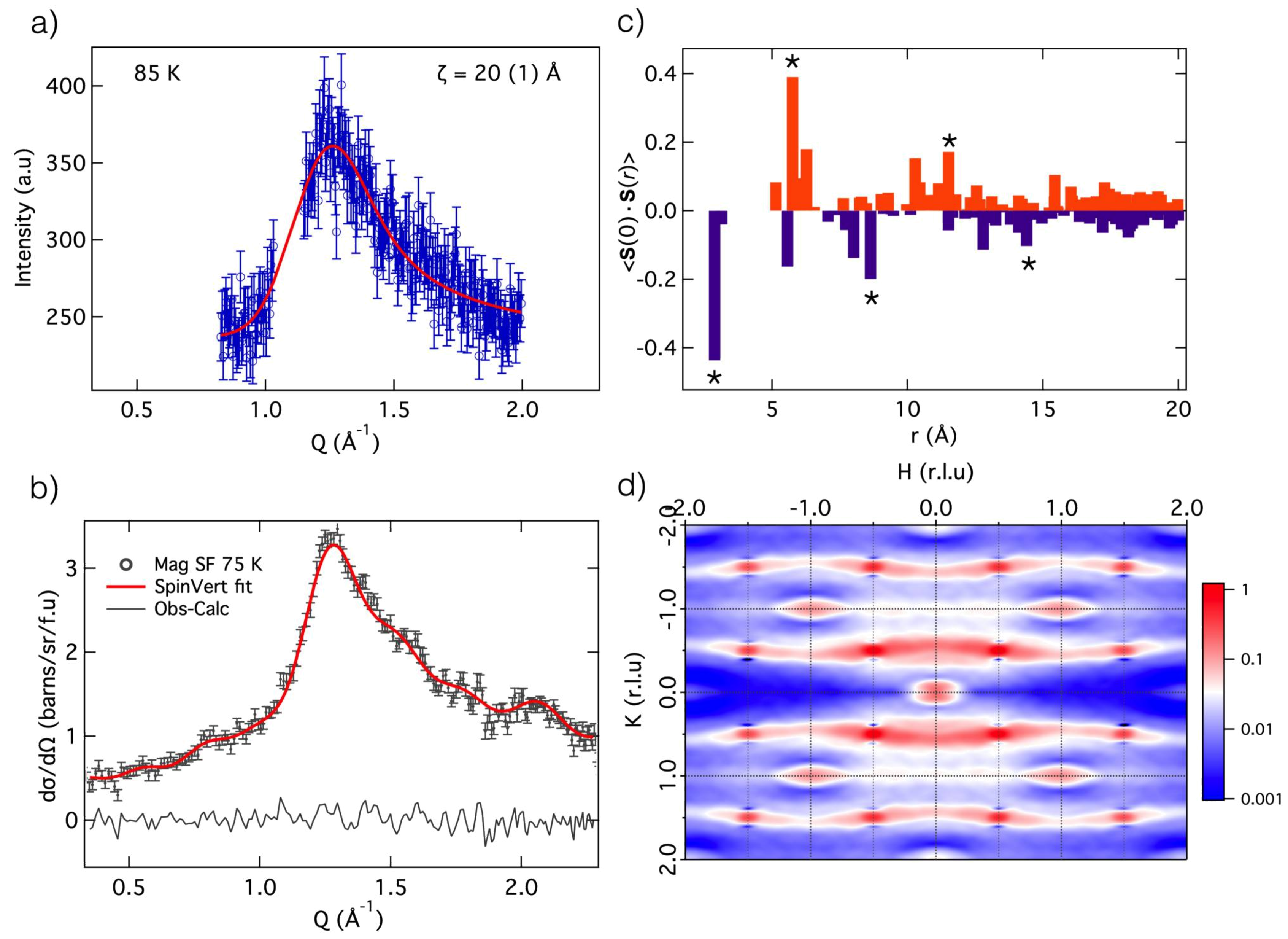}
\caption{(color online) a) Observed magnetic diffuse scattering for \cu~at 85 K using the E9 diffractometer. The (001) Bragg peak has been manually removed, and the solid line shows the result of fitting the Warren function (Eqn. 2); b) Observed quasi-elastic spin-flip magnetic neutron scattering signal recorded using D7 at 75 K. The solid line shows the result calculated using the SpinVert reverse Monte Carlo program; c) Extracted spin-spin correlations as a function of real space extracted from the fit in panel b). Negative (positive) values represent AFM (FM) correlations and those along the short $b$-axis direction are starred; d) Predicted single crystal elastic magnetic diffuse scattering for \cu~at 75K. }
\label{Fig5}
\end{center} 
\end{figure*}
Note that this issue is equally important for spallation neutron time-of-flight diffractometers used for e.g. mPDF experiments~\cite{frandsen2020nanoscale}, although harder to quantify. At the Q-range where magnetic diffuse scattering is strongest, the majority of neutrons which contribute may only have E$_{i}$ ~in the range of a few 10's of meV~\cite{Joerg}.\\
We now turn to our D7 measurements~\cite{data} in the paramagnetic regime. Judging by the inelastic data shown in Fig. 3, these are only sensitive to quasi-elastic magnetic correlations, and not to the gapped inelastic signal on the neutron energy loss (Stokes) side. As our experiment was performed at finite temperature, we also estimated the possible contribution of anti-Stokes scattered neutrons. This is necessary due to the different shape of the kinematic window on the energy gain side, which potentially accesses neutrons with $-10 < \Delta E <  0$ meV at Q=1.25 \AA$^{-1}$. However, the population factor for the energy corresponding to the maximum in the inelastic signal is only $\exp^{-\Delta E/k_{B}T}$=0.07 at 75 K. Furthermore, the efficiency of the D7 detectors drops markedly with increased neutron energy \cite{stewart2009disordered}. The real space results of our Monte Carlo fitting thus represent the infinite time spin-spin correlations, $G(\vec{r},\infty)$.\\
The observed spin-flip magnetic scattering for \cu~at 75 K is shown in Fig. 5b. Also shown is the calculated fit using the SpinVert program and the residual. The fit was performed using the parameters described in the experimental section. The results of this procedure were used to calculate the real space spin-spin correlations, $<\textbf{S}(0)\cdot \textbf{S}(r)>$, as shown in Fig. 5c as a function of distance. These are large for strong interactions, and their sign reflects their anti or ferromagnetic nature. Our key results from this section are as follows: 1) The strongest correlation (J$_{1}$~in Fig. 1) is, unsurprisingly, AFM and between nearest neighbours ($\sim$2.88 \AA) in the $b$-direction; 2) The next-nearest correlation along the $b$-axis ($\sim$5.76 \AA) is strongly FM as expected from the magnetic structure; 3) correlations between chains (J$_{2}$~in Fig. 1) are extremely weak AFM, and hence completely frustrated.\\
The correlation length of order along the $b$-axis can be seen by the further strong AFM correlation at 8.64 \AA. This represents the third-nearest neighbour in this direction. Based upon these results, we can therefore conclude with some confidence that \cu~has largely 1D spin correlations in the paramagnetic phase. Nearest-neighbour chains along the crystallographic $a$-axis are almost independent, although there is a weak tendency for next-nearest neighbour chains to be coupled in an AFM manner, as seen by the negative correlation at 5.57 \AA. Finally, we also calculated the predicted single crystal magnetic elastic diffuse scattering from our SpinVert fits. This is shown in Figure 5d, and reveals sheets of scattering running along the $[h0l]$ directions. These show a certain degree of corrugation, reflecting the weaker longer range interactions which couple chains together.\\
\section{\label{sec:level1}CONCLUSIONS}
In summary, we have performed detailed thermal and cold neutron scattering experiments on the frustrated two-dimensional triangular lattice material \cu. These reveal pronounced inelastic effects even in the paramagnetic state, which invalidate attempts to analyse diffuse scattering from thermal diffractometers. Examination of the quasi-elastic magnetic diffuse scattering using XYZ polarisation shows that the spin-spin correlations in the paramagnetic state are actually 1D, and not 2D as previously reported~\cite{garlea2011tuning,terada2011magnetic}. These results remove inconsistencies between neutron scattering results and physical property measurements, and establish similarities with related \na~in the paramagnetic phase. Our results strongly encourage the growth of single crystal samples suitable for inelastic neutron scattering.\\

We thank the Helmholtz-Zentrum Berlin and the ILL for access to instrumentation. Ce travail \'etait soutane par le programme "Investissements d'Avenir", projet ISITE-BFC (contrat ANR-15-IDEX-0003).
%%%%%%%
%\bibliographystyle{aipauth4-1}
\bibliography{D7}

\begin{thebibliography}{25}
\expandafter\ifx\csname natexlab\endcsname\relax\def\natexlab#1{#1}\fi
\expandafter\ifx\csname bibnamefont\endcsname\relax
  \def\bibnamefont#1{#1}\fi
\expandafter\ifx\csname bibfnamefont\endcsname\relax
  \def\bibfnamefont#1{#1}\fi
\expandafter\ifx\csname citenamefont\endcsname\relax
  \def\citenamefont#1{#1}\fi
\expandafter\ifx\csname url\endcsname\relax
  \def\url#1{\texttt{#1}}\fi
\expandafter\ifx\csname urlprefix\endcsname\relax\def\urlprefix{URL }\fi
\providecommand{\bibinfo}[2]{#2}
\providecommand{\eprint}[2][]{\url{#2}}

\bibitem[{\citenamefont{McQueen et~al.}(2008)\citenamefont{McQueen, Stephens,
  Huang, Klimczuk, Ronning, and Cava}}]{NaVO2}
\bibinfo{author}{\bibfnamefont{T.}~\bibnamefont{McQueen}},
  \bibinfo{author}{\bibfnamefont{P.}~\bibnamefont{Stephens}},
  \bibinfo{author}{\bibfnamefont{Q.}~\bibnamefont{Huang}},
  \bibinfo{author}{\bibfnamefont{T.}~\bibnamefont{Klimczuk}},
  \bibinfo{author}{\bibfnamefont{F.}~\bibnamefont{Ronning}}, \bibnamefont{and}
  \bibinfo{author}{\bibfnamefont{R.~J.} \bibnamefont{Cava}},
  \bibinfo{journal}{Physical review letters} \textbf{\bibinfo{volume}{101}},
  \bibinfo{pages}{166402} (\bibinfo{year}{2008}).

\bibitem[{\citenamefont{Frontzek et~al.}(2011)\citenamefont{Frontzek, Ehlers,
  Podlesnyak, Cao, Matsuda, Zaharko, Aliouane, Barilo, and
  Shiryaev}}]{frontzek2011magnetic}
\bibinfo{author}{\bibfnamefont{M.}~\bibnamefont{Frontzek}},
  \bibinfo{author}{\bibfnamefont{G.}~\bibnamefont{Ehlers}},
  \bibinfo{author}{\bibfnamefont{A.}~\bibnamefont{Podlesnyak}},
  \bibinfo{author}{\bibfnamefont{H.}~\bibnamefont{Cao}},
  \bibinfo{author}{\bibfnamefont{M.}~\bibnamefont{Matsuda}},
  \bibinfo{author}{\bibfnamefont{O.}~\bibnamefont{Zaharko}},
  \bibinfo{author}{\bibfnamefont{N.}~\bibnamefont{Aliouane}},
  \bibinfo{author}{\bibfnamefont{S.}~\bibnamefont{Barilo}}, \bibnamefont{and}
  \bibinfo{author}{\bibfnamefont{S.}~\bibnamefont{Shiryaev}},
  \bibinfo{journal}{Journal of Physics: Condensed Matter}
  \textbf{\bibinfo{volume}{24}}, \bibinfo{pages}{016004}
  (\bibinfo{year}{2011}).

\bibitem[{\citenamefont{Takatsu et~al.}(2009)\citenamefont{Takatsu, Yoshizawa,
  Yonezawa, and Maeno}}]{takatsu2009critical}
\bibinfo{author}{\bibfnamefont{H.}~\bibnamefont{Takatsu}},
  \bibinfo{author}{\bibfnamefont{H.}~\bibnamefont{Yoshizawa}},
  \bibinfo{author}{\bibfnamefont{S.}~\bibnamefont{Yonezawa}}, \bibnamefont{and}
  \bibinfo{author}{\bibfnamefont{Y.}~\bibnamefont{Maeno}},
  \bibinfo{journal}{Physical Review B} \textbf{\bibinfo{volume}{79}},
  \bibinfo{pages}{104424} (\bibinfo{year}{2009}).

\bibitem[{\citenamefont{Roger et~al.}(2007)\citenamefont{Roger, Morris,
  Tennant, Gutmann, Goff, Hoffmann, Feyerherm, Dudzik, Prabhakaran, Boothroyd
  et~al.}}]{roger2007patterning}
\bibinfo{author}{\bibfnamefont{M.}~\bibnamefont{Roger}},
  \bibinfo{author}{\bibfnamefont{D.}~\bibnamefont{Morris}},
  \bibinfo{author}{\bibfnamefont{D.}~\bibnamefont{Tennant}},
  \bibinfo{author}{\bibfnamefont{M.}~\bibnamefont{Gutmann}},
  \bibinfo{author}{\bibfnamefont{J.}~\bibnamefont{Goff}},
  \bibinfo{author}{\bibfnamefont{J.-U.} \bibnamefont{Hoffmann}},
  \bibinfo{author}{\bibfnamefont{R.}~\bibnamefont{Feyerherm}},
  \bibinfo{author}{\bibfnamefont{E.}~\bibnamefont{Dudzik}},
  \bibinfo{author}{\bibfnamefont{D.}~\bibnamefont{Prabhakaran}},
  \bibinfo{author}{\bibfnamefont{A.}~\bibnamefont{Boothroyd}},
  \bibnamefont{et~al.}, \bibinfo{journal}{Nature}
  \textbf{\bibinfo{volume}{445}}, \bibinfo{pages}{631} (\bibinfo{year}{2007}).

\bibitem[{\citenamefont{Mitsuda et~al.}(1998)\citenamefont{Mitsuda, Kasahara,
  Uno, and Mase}}]{mitsuda1998partially}
\bibinfo{author}{\bibfnamefont{S.}~\bibnamefont{Mitsuda}},
  \bibinfo{author}{\bibfnamefont{N.}~\bibnamefont{Kasahara}},
  \bibinfo{author}{\bibfnamefont{T.}~\bibnamefont{Uno}}, \bibnamefont{and}
  \bibinfo{author}{\bibfnamefont{M.}~\bibnamefont{Mase}},
  \bibinfo{journal}{Journal of the Physical Society of Japan}
  \textbf{\bibinfo{volume}{67}}, \bibinfo{pages}{4026} (\bibinfo{year}{1998}).

\bibitem[{\citenamefont{Ye et~al.}(2006)\citenamefont{Ye, Ren, Huang,
  Fernandez-Baca, Dai, Lynn, and Kimura}}]{ye2006spontaneous}
\bibinfo{author}{\bibfnamefont{F.}~\bibnamefont{Ye}},
  \bibinfo{author}{\bibfnamefont{Y.}~\bibnamefont{Ren}},
  \bibinfo{author}{\bibfnamefont{Q.}~\bibnamefont{Huang}},
  \bibinfo{author}{\bibfnamefont{J.~A.} \bibnamefont{Fernandez-Baca}},
  \bibinfo{author}{\bibfnamefont{P.}~\bibnamefont{Dai}},
  \bibinfo{author}{\bibfnamefont{J.}~\bibnamefont{Lynn}}, \bibnamefont{and}
  \bibinfo{author}{\bibfnamefont{T.}~\bibnamefont{Kimura}},
  \bibinfo{journal}{Physical Review B} \textbf{\bibinfo{volume}{73}},
  \bibinfo{pages}{220404} (\bibinfo{year}{2006}).

\bibitem[{\citenamefont{Giot et~al.}(2007)\citenamefont{Giot, Chapon,
  Androulakis, Green, Radaelli, and Lappas}}]{giot2007magnetoelastic}
\bibinfo{author}{\bibfnamefont{M.}~\bibnamefont{Giot}},
  \bibinfo{author}{\bibfnamefont{L.~C.} \bibnamefont{Chapon}},
  \bibinfo{author}{\bibfnamefont{J.}~\bibnamefont{Androulakis}},
  \bibinfo{author}{\bibfnamefont{M.~A.} \bibnamefont{Green}},
  \bibinfo{author}{\bibfnamefont{P.~G.} \bibnamefont{Radaelli}},
  \bibnamefont{and} \bibinfo{author}{\bibfnamefont{A.}~\bibnamefont{Lappas}},
  \bibinfo{journal}{Physical review letters} \textbf{\bibinfo{volume}{99}},
  \bibinfo{pages}{247211} (\bibinfo{year}{2007}).

\bibitem[{\citenamefont{Zorko et~al.}(2008)\citenamefont{Zorko, El~Shawish,
  Ar{\v{c}}on, Jagli{\v{c}}i{\'c}, Lappas, van Tol, and
  Brunel}}]{zorko2008magnetic}
\bibinfo{author}{\bibfnamefont{A.}~\bibnamefont{Zorko}},
  \bibinfo{author}{\bibfnamefont{S.}~\bibnamefont{El~Shawish}},
  \bibinfo{author}{\bibfnamefont{D.}~\bibnamefont{Ar{\v{c}}on}},
  \bibinfo{author}{\bibfnamefont{Z.}~\bibnamefont{Jagli{\v{c}}i{\'c}}},
  \bibinfo{author}{\bibfnamefont{A.}~\bibnamefont{Lappas}},
  \bibinfo{author}{\bibfnamefont{H.}~\bibnamefont{van Tol}}, \bibnamefont{and}
  \bibinfo{author}{\bibfnamefont{L.~C.} \bibnamefont{Brunel}},
  \bibinfo{journal}{Physical Review B} \textbf{\bibinfo{volume}{77}},
  \bibinfo{pages}{024412} (\bibinfo{year}{2008}).

\bibitem[{\citenamefont{Damay et~al.}(2009)\citenamefont{Damay, Poienar,
  Martin, Maignan, Rodriguez-Carvajal, Andr{\'e}, and Doumerc}}]{damay2009spin}
\bibinfo{author}{\bibfnamefont{F.}~\bibnamefont{Damay}},
  \bibinfo{author}{\bibfnamefont{M.}~\bibnamefont{Poienar}},
  \bibinfo{author}{\bibfnamefont{C.}~\bibnamefont{Martin}},
  \bibinfo{author}{\bibfnamefont{A.}~\bibnamefont{Maignan}},
  \bibinfo{author}{\bibfnamefont{J.}~\bibnamefont{Rodriguez-Carvajal}},
  \bibinfo{author}{\bibfnamefont{G.}~\bibnamefont{Andr{\'e}}},
  \bibnamefont{and} \bibinfo{author}{\bibfnamefont{J.-P.}
  \bibnamefont{Doumerc}}, \bibinfo{journal}{Physical Review B}
  \textbf{\bibinfo{volume}{80}}, \bibinfo{pages}{094410}
  (\bibinfo{year}{2009}).

\bibitem[{\citenamefont{Coldea et~al.}(2003)\citenamefont{Coldea, Tennant, and
  Tylczynski}}]{coldea2003extended}
\bibinfo{author}{\bibfnamefont{R.}~\bibnamefont{Coldea}},
  \bibinfo{author}{\bibfnamefont{D.}~\bibnamefont{Tennant}}, \bibnamefont{and}
  \bibinfo{author}{\bibfnamefont{Z.}~\bibnamefont{Tylczynski}},
  \bibinfo{journal}{Physical Review B} \textbf{\bibinfo{volume}{68}},
  \bibinfo{pages}{134424} (\bibinfo{year}{2003}).

\bibitem[{\citenamefont{Stock et~al.}(2009)\citenamefont{Stock, Chapon,
  Adamopoulos, Lappas, Giot, Taylor, Green, Brown, and
  Radaelli}}]{stock2009one}
\bibinfo{author}{\bibfnamefont{C.}~\bibnamefont{Stock}},
  \bibinfo{author}{\bibfnamefont{L.}~\bibnamefont{Chapon}},
  \bibinfo{author}{\bibfnamefont{O.}~\bibnamefont{Adamopoulos}},
  \bibinfo{author}{\bibfnamefont{A.}~\bibnamefont{Lappas}},
  \bibinfo{author}{\bibfnamefont{M.}~\bibnamefont{Giot}},
  \bibinfo{author}{\bibfnamefont{J.}~\bibnamefont{Taylor}},
  \bibinfo{author}{\bibfnamefont{M.}~\bibnamefont{Green}},
  \bibinfo{author}{\bibfnamefont{C.}~\bibnamefont{Brown}}, \bibnamefont{and}
  \bibinfo{author}{\bibfnamefont{P.}~\bibnamefont{Radaelli}},
  \bibinfo{journal}{Physical review letters} \textbf{\bibinfo{volume}{103}},
  \bibinfo{pages}{077202} (\bibinfo{year}{2009}).

\bibitem[{\citenamefont{Dally et~al.}(2018)\citenamefont{Dally, Zhao, Xu,
  Chisnell, Stone, Lynn, Balents, and Wilson}}]{dally2018amplitude}
\bibinfo{author}{\bibfnamefont{R.~L.} \bibnamefont{Dally}},
  \bibinfo{author}{\bibfnamefont{Y.}~\bibnamefont{Zhao}},
  \bibinfo{author}{\bibfnamefont{Z.}~\bibnamefont{Xu}},
  \bibinfo{author}{\bibfnamefont{R.}~\bibnamefont{Chisnell}},
  \bibinfo{author}{\bibfnamefont{M.~B.} \bibnamefont{Stone}},
  \bibinfo{author}{\bibfnamefont{J.~W.} \bibnamefont{Lynn}},
  \bibinfo{author}{\bibfnamefont{L.}~\bibnamefont{Balents}}, \bibnamefont{and}
  \bibinfo{author}{\bibfnamefont{S.~D.} \bibnamefont{Wilson}},
  \bibinfo{journal}{Nature communications} \textbf{\bibinfo{volume}{9}},
  \bibinfo{pages}{1} (\bibinfo{year}{2018}).

\bibitem[{\citenamefont{Garlea et~al.}(2011)\citenamefont{Garlea, Savici, and
  Jin}}]{garlea2011tuning}
\bibinfo{author}{\bibfnamefont{V.~O.} \bibnamefont{Garlea}},
  \bibinfo{author}{\bibfnamefont{A.~T.} \bibnamefont{Savici}},
  \bibnamefont{and} \bibinfo{author}{\bibfnamefont{R.}~\bibnamefont{Jin}},
  \bibinfo{journal}{Physical Review B} \textbf{\bibinfo{volume}{83}},
  \bibinfo{pages}{172407} (\bibinfo{year}{2011}).

\bibitem[{\citenamefont{Terada et~al.}(2011)\citenamefont{Terada, Tsuchiya,
  Kitazawa, Osakabe, Metoki, Igawa, and Ohoyama}}]{terada2011magnetic}
\bibinfo{author}{\bibfnamefont{N.}~\bibnamefont{Terada}},
  \bibinfo{author}{\bibfnamefont{Y.}~\bibnamefont{Tsuchiya}},
  \bibinfo{author}{\bibfnamefont{H.}~\bibnamefont{Kitazawa}},
  \bibinfo{author}{\bibfnamefont{T.}~\bibnamefont{Osakabe}},
  \bibinfo{author}{\bibfnamefont{N.}~\bibnamefont{Metoki}},
  \bibinfo{author}{\bibfnamefont{N.}~\bibnamefont{Igawa}}, \bibnamefont{and}
  \bibinfo{author}{\bibfnamefont{K.}~\bibnamefont{Ohoyama}},
  \bibinfo{journal}{Physical Review B} \textbf{\bibinfo{volume}{84}},
  \bibinfo{pages}{064432} (\bibinfo{year}{2011}).

\bibitem[{\citenamefont{Frandsen et~al.}(2020)\citenamefont{Frandsen, Bozin,
  Aza, Mart{\'\i}nez, Feygenson, Page, and Lappas}}]{frandsen2020nanoscale}
\bibinfo{author}{\bibfnamefont{B.~A.} \bibnamefont{Frandsen}},
  \bibinfo{author}{\bibfnamefont{E.~S.} \bibnamefont{Bozin}},
  \bibinfo{author}{\bibfnamefont{E.}~\bibnamefont{Aza}},
  \bibinfo{author}{\bibfnamefont{A.~F.} \bibnamefont{Mart{\'\i}nez}},
  \bibinfo{author}{\bibfnamefont{M.}~\bibnamefont{Feygenson}},
  \bibinfo{author}{\bibfnamefont{K.}~\bibnamefont{Page}}, \bibnamefont{and}
  \bibinfo{author}{\bibfnamefont{A.}~\bibnamefont{Lappas}},
  \bibinfo{journal}{Physical Review B} \textbf{\bibinfo{volume}{101}},
  \bibinfo{pages}{024423} (\bibinfo{year}{2020}).

\bibitem[{\citenamefont{Momma and Izumi}(2011)}]{momma2011vesta}
\bibinfo{author}{\bibfnamefont{K.}~\bibnamefont{Momma}} \bibnamefont{and}
  \bibinfo{author}{\bibfnamefont{F.}~\bibnamefont{Izumi}},
  \bibinfo{journal}{Journal of applied crystallography}
  \textbf{\bibinfo{volume}{44}}, \bibinfo{pages}{1272} (\bibinfo{year}{2011}).

\bibitem[{\citenamefont{Stewart et~al.}(2009)\citenamefont{Stewart, Deen,
  Andersen, Schober, Barth{\'e}l{\'e}my, Hillier, Murani, Hayes, and
  Lindenau}}]{stewart2009disordered}
\bibinfo{author}{\bibfnamefont{J.}~\bibnamefont{Stewart}},
  \bibinfo{author}{\bibfnamefont{P.}~\bibnamefont{Deen}},
  \bibinfo{author}{\bibfnamefont{K.}~\bibnamefont{Andersen}},
  \bibinfo{author}{\bibfnamefont{H.}~\bibnamefont{Schober}},
  \bibinfo{author}{\bibfnamefont{J.-F.} \bibnamefont{Barth{\'e}l{\'e}my}},
  \bibinfo{author}{\bibfnamefont{J.}~\bibnamefont{Hillier}},
  \bibinfo{author}{\bibfnamefont{A.}~\bibnamefont{Murani}},
  \bibinfo{author}{\bibfnamefont{T.}~\bibnamefont{Hayes}}, \bibnamefont{and}
  \bibinfo{author}{\bibfnamefont{B.}~\bibnamefont{Lindenau}},
  \bibinfo{journal}{Journal of Applied Crystallography}
  \textbf{\bibinfo{volume}{42}}, \bibinfo{pages}{69} (\bibinfo{year}{2009}).

\bibitem[{\citenamefont{Fennell et~al.}(2017)\citenamefont{Fennell,
  Mangin-Thro, Mutka, Nilsen, and Wildes}}]{fennell2017wavevector}
\bibinfo{author}{\bibfnamefont{T.}~\bibnamefont{Fennell}},
  \bibinfo{author}{\bibfnamefont{L.}~\bibnamefont{Mangin-Thro}},
  \bibinfo{author}{\bibfnamefont{H.}~\bibnamefont{Mutka}},
  \bibinfo{author}{\bibfnamefont{G.}~\bibnamefont{Nilsen}}, \bibnamefont{and}
  \bibinfo{author}{\bibfnamefont{A.}~\bibnamefont{Wildes}},
  \bibinfo{journal}{Nuclear Instruments and Methods in Physics Research Section
  A: Accelerators, Spectrometers, Detectors and Associated Equipment}
  \textbf{\bibinfo{volume}{857}}, \bibinfo{pages}{24} (\bibinfo{year}{2017}).

\bibitem[{\citenamefont{Ehlers et~al.}(2013)\citenamefont{Ehlers, Stewart,
  Wildes, Deen, and Andersen}}]{ehlers2013generalization}
\bibinfo{author}{\bibfnamefont{G.}~\bibnamefont{Ehlers}},
  \bibinfo{author}{\bibfnamefont{J.~R.} \bibnamefont{Stewart}},
  \bibinfo{author}{\bibfnamefont{A.}~\bibnamefont{Wildes}},
  \bibinfo{author}{\bibfnamefont{P.}~\bibnamefont{Deen}}, \bibnamefont{and}
  \bibinfo{author}{\bibfnamefont{K.}~\bibnamefont{Andersen}},
  \bibinfo{journal}{Review of Scientific Instruments}
  \textbf{\bibinfo{volume}{84}}, \bibinfo{pages}{093901}
  (\bibinfo{year}{2013}).

\bibitem[{\citenamefont{Paddison et~al.}(2013)\citenamefont{Paddison, Stewart,
  and Goodwin}}]{paddison2013spinvert}
\bibinfo{author}{\bibfnamefont{J.~A.} \bibnamefont{Paddison}},
  \bibinfo{author}{\bibfnamefont{J.~R.} \bibnamefont{Stewart}},
  \bibnamefont{and} \bibinfo{author}{\bibfnamefont{A.~L.}
  \bibnamefont{Goodwin}}, \bibinfo{journal}{Journal of Physics: Condensed
  Matter} \textbf{\bibinfo{volume}{25}}, \bibinfo{pages}{454220}
  (\bibinfo{year}{2013}).

\bibitem[{\citenamefont{F{\aa}k and Dorner}(1997)}]{faak1997phonon}
\bibinfo{author}{\bibfnamefont{B.}~\bibnamefont{F{\aa}k}} \bibnamefont{and}
  \bibinfo{author}{\bibfnamefont{B.}~\bibnamefont{Dorner}},
  \bibinfo{journal}{Physica B: Condensed Matter}
  \textbf{\bibinfo{volume}{234}}, \bibinfo{pages}{1107} (\bibinfo{year}{1997}).

\bibitem[{\citenamefont{Krimmel et~al.}(2009)\citenamefont{Krimmel, Mutka,
  Koza, Tsurkan, and Loidl}}]{krimmel2009spin}
\bibinfo{author}{\bibfnamefont{A.}~\bibnamefont{Krimmel}},
  \bibinfo{author}{\bibfnamefont{H.}~\bibnamefont{Mutka}},
  \bibinfo{author}{\bibfnamefont{M.}~\bibnamefont{Koza}},
  \bibinfo{author}{\bibfnamefont{V.}~\bibnamefont{Tsurkan}}, \bibnamefont{and}
  \bibinfo{author}{\bibfnamefont{A.}~\bibnamefont{Loidl}},
  \bibinfo{journal}{Physical Review B} \textbf{\bibinfo{volume}{79}},
  \bibinfo{pages}{134406} (\bibinfo{year}{2009}).

\bibitem[{\citenamefont{Warren}(1941)}]{warren1941x}
\bibinfo{author}{\bibfnamefont{B.}~\bibnamefont{Warren}},
  \bibinfo{journal}{Physical Review} \textbf{\bibinfo{volume}{59}},
  \bibinfo{pages}{693} (\bibinfo{year}{1941}).

\bibitem[{\citenamefont{Neuefeind}()}]{Joerg}
\bibinfo{author}{\bibfnamefont{J.}~\bibnamefont{Neuefeind}},
  \bibinfo{howpublished}{personal communication}.

\bibitem[{\citenamefont{ILL}(2015)}]{data}
\bibinfo{author}{\bibnamefont{ILL}},
  \bibinfo{journal}{DOI:10.529/ILL-DATA-TEST-2587}  (\bibinfo{year}{2015}).

\end{thebibliography}

%%%%%%%
\end{document}